\documentclass[aps,prx,preprint,groupedaddress,showpacs]{revtex4-2}

\usepackage{graphicx}
\usepackage{amssymb}
\usepackage{color} 
\usepackage{epstopdf}
\usepackage{soul}

\begin{document}


\title{Impact craters formed by spinning granular projectiles\\
	\textnormal{Accepted manuscript for Physical Review E, 108, 054904, (2023), DOI: 10.1103/PhysRevE.108.054904, https://doi.org/10.1103/PhysRevE.108.054904}}



\author{Douglas D. Carvalho}
\author{Nicolao C. Lima}
\author{Erick M. Franklin}
 \email{erick.franklin@unicamp.br}
 \thanks{Corresponding author}
\affiliation{%
School of Mechanical Engineering, UNICAMP - University of Campinas,\\
Rua Mendeleyev, 200, Campinas, SP, Brazil\\
}%


\date{\today}

\begin{abstract}
Craters formed by the impact of agglomerated materials are commonly observed in nature, such as asteroids colliding with planets and moons. In this paper, we investigate how the projectile spin and cohesion lead to different crater shapes. For that, we carried out DEM (discrete element method) computations of spinning granular projectiles impacting onto cohesionless grains, for different bonding stresses, initial spins and initial heights. We found that, as the bonding stresses decrease and the initial spin increases, the projectile's grains spread farther from the collision point, and, in consequence, the crater shape becomes flatter, with peaks around the rim and in the center of craters. Our results shed light on the dispersion of the projectile's material and the different shapes of craters found on Earth and other planetary environments.
\end{abstract}


\maketitle


\section{INTRODUCTION}
\label{sec:intro}

Craters formed by the impact of projectiles are commonly observed in nature, such as km-size asteroids colliding with planets and moons and cm-size seeds falling from trees. While the latter involves very low energies (as low as 10$^{-7}$ J, the equivalent of lighting a LED lamp for approximately 0.0000001 s), the former involves huge energies that surpass that of a hydrogen bomb (from 10$^{16}$ J on). Because those scales differ by more than 23 orders of magnitude, the cratering processes and resulting shapes are not the same in all cases. For example, under small energies (low masses and velocities) the impact results in the partial penetration of the projectile and ejection of ground material, while under large energies it involves also melting and evaporation.

\begin{figure}[h!]
	\begin{center}
		\includegraphics[width=0.95\linewidth]{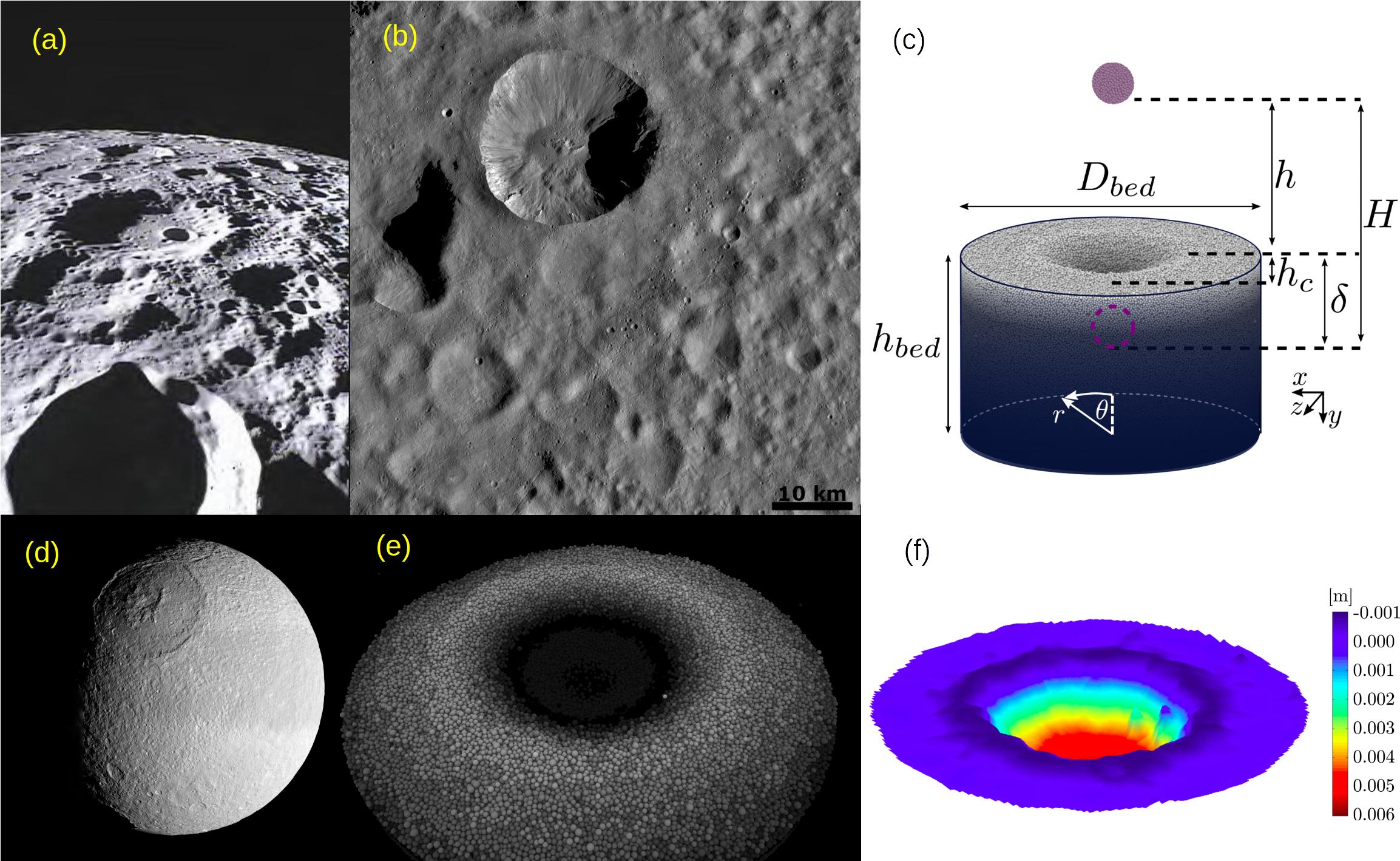}\\
	\end{center}
	\caption{(a) Craters on Earth's moon (in the middle, with a smaller bow-shaped crater inside, is Poinsot crater); (b) Craters on Vesta, with a recent 20-km-diameter crater on the top of image; (c) layout of the numerical setup (the $y$ coordinate points downwards, and, although shown on the bottom, the origin of the coordinate system is on the bed surface centered horizontally in the domain); (d)  445-km-diameter crater on Saturn's moon Tethys; (e) 76-mm-diameter crater obtained numerically from the impact of a 25-mm-diameter steel sphere falling from 50 mm onto a bed of particles (glass spheres with mean diameter of 1 mm); (f) topography (elevation) of a crater formed by a spinning projectile consisting of bonded grains (we notice at least one internal peak close to the rim). In this figure, the bonding stresses are 10$^7$ N/m$^2$, the ratio between linear and angular kinetic energies is 1, and the colorbar shows the elevation from the undisturbed surface (pointing downwards). Images in panels (a), (b) and (d): Courtesy NASA/JPL-Caltech.}
	\label{fig:craters}
\end{figure}

Craters of distinct shape and size have been observed in environments with different ground properties and gravity acceleration \cite{Melosh}, so that strong variations occur and a classification is not straightforward \cite{Barlow, Arvidson}. In general, small craters have a bowl shape (also called simple craters, Fig. \ref{fig:craters}(b)), and, as the craters become larger, they present a flat floor and a central peak or peak rings. For even larger craters, they have, in addition to the flat floor and central peak (or peak rings), external rings that are formed from the partial collapse of steep walls (Fig. \ref{fig:craters}(d)). For reference, lunar craters with diameters smaller than approximately 10 km are bowl-shaped, those with diameters of the order of 100 km have the external rings, flat floor and peak rings, and craters within those values vary between bowl-shaped and flat floor with central peak \cite{Melosh}. Notwithstanding their ubiquitous nature, the mechanisms leading to different crater shapes are far from being completely understood.

Besides the involved energies and sizes, other properties such as the projectile and ground compositions \cite{Pacheco2}, confinement \cite{Seguin2}, projectile spin \cite{Carvalho}, and microscopic friction \cite{Carvalho} can strongly influence the crater shape. By using dimensional analysis, Holsapple \cite{Holsapple} showed that the dimensionless volume of the crater is a function of two pressure ratios and the density ratio $\rho_p/\rho$, where $\rho_p$ and $\rho$ are the densities of the projectile and ground materials, respectively. One of the pressure ratios consists of the projectile weight divided by its surface area and normalized by the dynamic pressure,

\begin{equation}
	\mathrm{Fr}^{-1} = \frac{D_p g}{V_p^2} \, ,
	\label{Eq:P2}
\end{equation}

\noindent where $D_p$ is the projectile diameter, $V_p$ is the velocity of the projectile at the impact, and $g$ is the modulus of gravity acceleration $\vec{g}$. This pressure ratio is the equivalent of the inverse of a Froude number Fr$^{-1}$ (gravitational effects compared to inertia), being important in geophysical processes, for which 10$^{-6}$ $\lesssim$ Fr$^{-1}$ $\lesssim$ 10$^{-2}$. Impact cratering is usually considered in the so-called gravity regime when Fr$^{-1}$ $\lesssim$ 10$^{-2}$, but for cohesionless grains the upper limit is acknowledged to be greater \cite{Holsapple}.

Because impacts of km-scale asteroids are rare events within the human timescale (every millions of years on Earth's surface, for example), laboratory-scale experiments and numerical simulations have proven essential in the investigation of crater formation \cite{Uehara, Walsh, Ciamarra, Katsuragi, deVet, Goldman, Seguin2, Seguin, Umbanhowar, Katsuragi2}. By assuring Fr$^{-1}$ $\lesssim$ 10$^{-1}$ in most cases and using targets consisting of cohesionless grains, those works allowed extrapolations of laboratory results to geophysical problems \cite{Suarez}. For example, Uehara et al. \cite{Uehara, Uehara2} carried out experiments where solid spheres were let to fall onto cohesionless grains for different $\rho_p$ and heights from the bed $h$, resulting in partially penetrating projectiles (penetration depth $\delta$ = crater depth $h_c$). They found that the crater diameter $D_c$ varies with $\left(\rho_p D_p^3 H \right)^{1/4}$, so that $D_c$ $\sim$ $E^{1/4}$ (the crater diameter varies as a 1/4 power of the energy), where $E$ is the available energy at the impact and $H$ = $h + \delta$ is the total drop distance. They also found that the crater depth $h_c$ does not scale with $E$, but $h_c \sim H^{1/3}$. In addition, they showed that the diameter of grains and friction and restitution coefficients of the projectile do not affect the crater diameter $D_c$.

Although most of experiments on impact cratering were for solid projectiles, many problems, in particular in geophysics, concern the impact of aggregates. For example, in the case of asteroids or meteors impacting the surface of a planet, aggregates can be divided into smaller parts which, in their turn, penetrate into the target and excavate the crater. This process can be responsible for the spreading of materials on Earth, just below the ground surface, such as nickel, platinum and gold \cite{Ganapathy, Sawlowicz, McDonald}. The impact of non-spinning aggregates was inquired into by Pacheco-V\'azquez and Ruiz-Su\'arez, who investigated first the sinking of collections of a few intruders in a low-density granular medium \cite{Pacheco} and afterwards the impact of aggregates onto a granular bed \cite{Pacheco2}. They showed that the same scale $D_c \sim h^{1/4}$ found for solid projectiles remains valid, but $D_c$ is larger for aggregates, with a discontinuity accounting for the energy necessary for fragmentation. As a consequence, complex crater shapes that depend on the packing fraction of the projectile appear. They also showed that $h_c \sim h^{1/3}$ is valid only for small energies: $h_c$ decreases abruptly above a threshold value and remains constant for higher energies. Finally, if the fragments once forming the projectile sink in the granular bed (which can happen in low-density beds), they move with a cooperative dynamics \cite{Pacheco}.

Recently, we \cite{Carvalho} carried out 3D (three dimensional) DEM (discrete element method) simulations and showed that the microscopic friction affects considerably the crater morphology. In addition, we showed that differences in initial packing fractions can engender the diversity of scaling laws found in the literature \cite{Uehara, Uehara2, Walsh, Seguin2, Katsuragi2}, and proposed an \textit{ad hoc} scaling that collapsed our data for the penetration length and can perhaps unify the existing correlations. Finally, we investigated the initial spin of the projectile and showed that both $\delta$ and $D_c$ increase with the projectile spin, that large asymmetries can apear depending on the spin axis, and that the final rebound of the projectile is suppressed by the spin.

Even though previous studies explained important aspects of impact cratering, many questions remain open. One of them concerns the mechanics of cratering for spinning aggregates impacting a granular ground. In this specific case, close to impacts observed in nature, the total or partial collapse of projectiles can engender different crater structures, explaining some of the crater shapes observed in nature and how materials from the projectile spread below and over the ground. This paper inquires into these questions. For that, we carried out 3D DEM computations of spinning granular projectiles (aggregates) impacting onto a bed consisting of cohesionless grains, for different bonding stresses (between the projectile's grains), initial spins and initial heights. We show that, as the bonding stresses decrease and the initial spin increases, the projectile's grains spread farther from the collision point, and, in consequence, the crater shape becomes flatter, with peaks around the rim and in the center of craters. In addition, we found that the penetration depth of rotating projectiles varies with their angular velocity and degree of collapse (number of detached particles), but not necessarily with the bonding stresses, indicating that under high spinning velocities the excess of breaking energy contributes only for the larger spreading in the horizontal plane and formation of peaks. Our results shed light on the different shapes of craters found on planets and moons, as well as on the distribution of the projectile material below and over the ground.

\section{BASIC EQUATIONS AND NUMERICAL SETUP}
\label{sec:mumerical}

As in Ref. \cite{Carvalho}, we carried out 3D DEM computations \cite{Cundall} using the open-source code LIGGGHTS \cite{Kloss2, Berger}. The code solves the linear (Eq. (\ref{Fp})) and angular (Eq. (\ref{Tp})) momentum equations for each individual particle at each time step,

\begin{equation}
	m\frac{d\vec{u}}{dt}= \vec{F}_{c} + m\vec{g} \,\, ,
	\label{Fp}
\end{equation}

\begin{equation}
	I\frac{d\vec{\omega}}{dt}=\vec{T}_{c} \,\, .
	\label{Tp}
\end{equation}

\noindent For each particle, $m$ is the mass, $\vec{u}$ is the velocity, $I$ is the moment of inertia, $\vec{\omega}$ is the angular velocity, $\vec{F}_{c}$ is the resultant of contact forces between solids, and $\vec{T}_{c}$ is the resultant of contact torques between solids. The contact forces and torques are computed using the elastic Hertz-Mindlin contact model \cite{direnzo}, and we take into account the rolling resistance (please see the Supplemental Material \cite{Supplemental} or Ref. \cite{Carvalho} for the model description).

The numerical domain consisted of: (i) $N$ $\sim$ 10$^6$ spheres with diameter 0.6 mm $\leq$ $d$ $\leq$ 1.4 mm following a Gaussian distribution and fixed density $\rho$ = 2600 kg/m$^3$, which formed a granular bed in a cylindrical container (the distribution of diameters used in the simulations is shown in the Supplemental Material \cite{Supplemental}); and (ii) $N_p$ = 1710 spheres with $d_p$ = 1 mm and $\rho_p$ = 15523 kg/m$^3$ bonded together, which formed a round projectile with total diameter $D_p$ = 0.015 m and bulk density $\rho_{p,bulk}$ = 7865 kg/m$^3$ (packing fraction $\phi_p$ = 0.507). Prior to each simulation, around 10$^6$ grains (bed spheres) were let to fall freely and settle, and grains that were above that height were deleted in order to have a horizontal surface (around 10$^4$ grains were removed), the number $N$ then depending on the initialization (being always $\sim$ 10$^6$). With that, we obtained a granular bed with diameter $D_{bed}$ = 125 mm, height $h_{bed}$ = 76.5 mm, and packing fraction $\phi$ = 0.554. For the projectile, the value of $\rho_p$ assured that the agglomerated material had the same size and mass of solid projectiles investigated in \cite{Carvalho}, and we applied a given bonding stress $\sigma_p$ to all grain-grain contacts. In our simulations, $\sigma_p$ was modeled through a breakup-tension threshold, and we used either $\sigma_p$ = $1 \times 10^{7}$, $5 \times 10^{7}$ or $1 \times 10^{32}$ N/m$^2$ in order to investigate the effect of bonding stresses on cratering. The highest value was chosen to avoid the projectile collapse, and the others to have partial or total collapses. The material that bonds two or more particles together can be modeled in several ways \cite{Guo, Schramm, Chen, Gong}. In this work, it acts as a spring and damper system, where the bonds can twist, bend, stretch and break due to both normal and tangential stresses. The damping system is based on Guo’s model \cite{Guo}, whereas the bond normal force and the bending and torsional moments are determined using linear models. More details are available in the Supplemental Material \cite{Supplemental}, and validation and details of the used model can be found in Guo et al. \cite{Guo} and Schramm et al. \cite{Schramm}.

\begin{table}[!h]
	\centering
	\caption{Properties of materials used in the simulations: $E$ is Young's modulus, $\nu$ is the Poisson ratio, and $\rho$ is the material density. The last column corresponds to the diameter of the considered object.}
	\label{tabmaterials}
	\begin{tabular}{l|c|c|c|c|c}
		\hline
		& \textbf{Material} & \textbf{$E$ (Pa)} & \textbf{$\nu$} & \textbf{$\rho$ (kg/m$^{3}$)}& \textbf{Diameters (mm)}\\
		\hline		
		Bed grains & Sand\footnotesize{$^{(1)-(2)}$} & $0.1 \times 10^{9}$ & 0.3 & 2600 & 0.6 $\leq$ $d$ $\leq$ 1.4 \\
		
		Projectile grains & - & $0.2 \times 10^{11}$ & 0.3 & 15523 & 1.0 \\
		
		Bond material & - & $0.2 \times 10^{11}$ & 0.3 & - & 0.1 \\
		
		Walls & Steel \footnotesize{$^{(1)}$} & $0.2 \times 10^{12}$ & 0.3 & 7865 & 125\\    
		   
		\hline
		\multicolumn{3}{l}{\footnotesize{$^{(1)}$ Ucgul et al. \cite{Ucgul1, Ucgul2, Ucgul3}}} \\
		\multicolumn{3}{l}{\footnotesize{$^{(2)}$ Derakhshani et al. \cite{Derakhshani}}}\\		
	\end{tabular}
\end{table}

\begin{table}[!h]
	\centering
	\caption{Coefficients used in the numerical simulations.}
	\label{tabcoefficients}
	\begin{tabular}{l|c|c}
		\hline
		\textbf{Coefficient}  & \textbf{Symbol} & \textbf{Value} \\
		\hline		  
		Restitution coefficient (bed grain-bed grain)\footnotesize{$^{(1)}$} & $\epsilon_{gg}$ & 0.60 \\
		Restitution coefficient (bed grain-projectile grain)\footnotesize{$^{(1)}$} & $\epsilon_{gp}$ & 0.60 \\
		Restitution coefficient (projectile grain-projectile grain)\footnotesize{$^{(2)}$} & $\epsilon_{pp}$ & 0.56\\
		Restitution coefficient (bed grain-wall)\footnotesize{$^{(1)}$} & $\epsilon_{gw}$ & 0.60 \\
		Restitution coefficient (projectile grain-wall)\footnotesize{$^{(1)}$} & $\epsilon_{pw}$ & 0.60 \\
		Fiction coefficient (bed grain-bed grain)\footnotesize{$^{(1),(3)}$} & $\mu_{gg}$ & 0.52 \\
		Friction coefficient (bed grain-projectile grain)\footnotesize{$^{(1)}$} & $\mu_{gp}$ & 0.50 \\
		Friction coefficient (projectile grain-projectile grain) & $\mu_{pp}$ & 0.57 \\
		Friction coefficient (bed grain-wall)\footnotesize{$^{(1)}$} & $\mu_{gw}$ & 0.50 \\
		Friction coefficient (projectile grain-wall) & $\mu_{pw}$ & 1.00 \\		
		Coefficient of rolling friction (bed grain-bed grain)\footnotesize{$^{(3)}$} & $\mu_{r,gg}$ & 0.30\\
		Coefficient of rolling friction (bed grain-projectile grain)\footnotesize{$^{(1)}$} & $\mu_{r,gp}$ & 0.05\\
		Coefficient of rolling friction (projectile grain-projectile grain) & $\mu_{r,pp}$ & 0.30\\
		Coefficient of rolling friction (bed grain-wall)\footnotesize{$^{(1)}$} & $\mu_{r,gw}$ & 0.05\\
		Coefficient of rolling friction (projectile grain-wall) & $\mu_{r,pw}$ & 1.00\\
		\hline
		\multicolumn{3}{l}{\footnotesize{$^{(1)}$ Ucgul et al. \cite{Ucgul1, Ucgul2, Ucgul3}}} \\
		\multicolumn{3}{l}{\footnotesize{$^{(2)}$ Zaikin et al. \cite{Zaikin}}} \\
		\multicolumn{3}{l}{\footnotesize{$^{(3)}$ Derakhshani et al. \cite{Derakhshani}}}
	\end{tabular}
\end{table}

The properties and coefficients of grains forming the bed and projectile were taken from the literature, and are listed in Tabs. \ref{tabmaterials} and \ref{tabcoefficients} (together with those for the walls). In addition, we validated the friction coefficients listed in Tab. \ref{tabcoefficients} by measuring the angles of repose obtained numerically (details available in Ref. \cite{Carvalho}). Because we used spherical particles, we embedded angularity in the rolling friction $\mu_r$ (for typical sand, Derakhshani et al. \cite{Derakhshani} showed that  $\mu_r$ = 0.3). The simulations began by imposing to the projectile a collision velocity $V_p$ corresponding to the free-fall height $h$, i.e., $V_p$ = $\sqrt{2gh}$. For the values used in our simulations, Froude numbers were within 3.8 $\times$ 10$^{-3}$ $\leq$ Fr$^{-1}$ $\leq$ 7.5 $\times$ 10$^{-2}$, and we used a time step $\Delta t = 1 \times 10^{-7}$ s, which corresponds to less than 10 \% of the Rayleigh time \cite{Derakhshani}. Figure \ref{fig:craters}(c) shows a layout of the numerical setup, and animations showing impacts and cratering are available in the Supplemental Material \cite{Supplemental}. The numerical setup of our simulations, output files, and scripts for post-processing the outputs are available in an open repository \cite{Supplemental2}.

\section{\label{sec:Results} RESULTS AND DISCUSSION}
  
\begin{figure}[ht]
	\begin{center}
		\includegraphics[width=\linewidth]{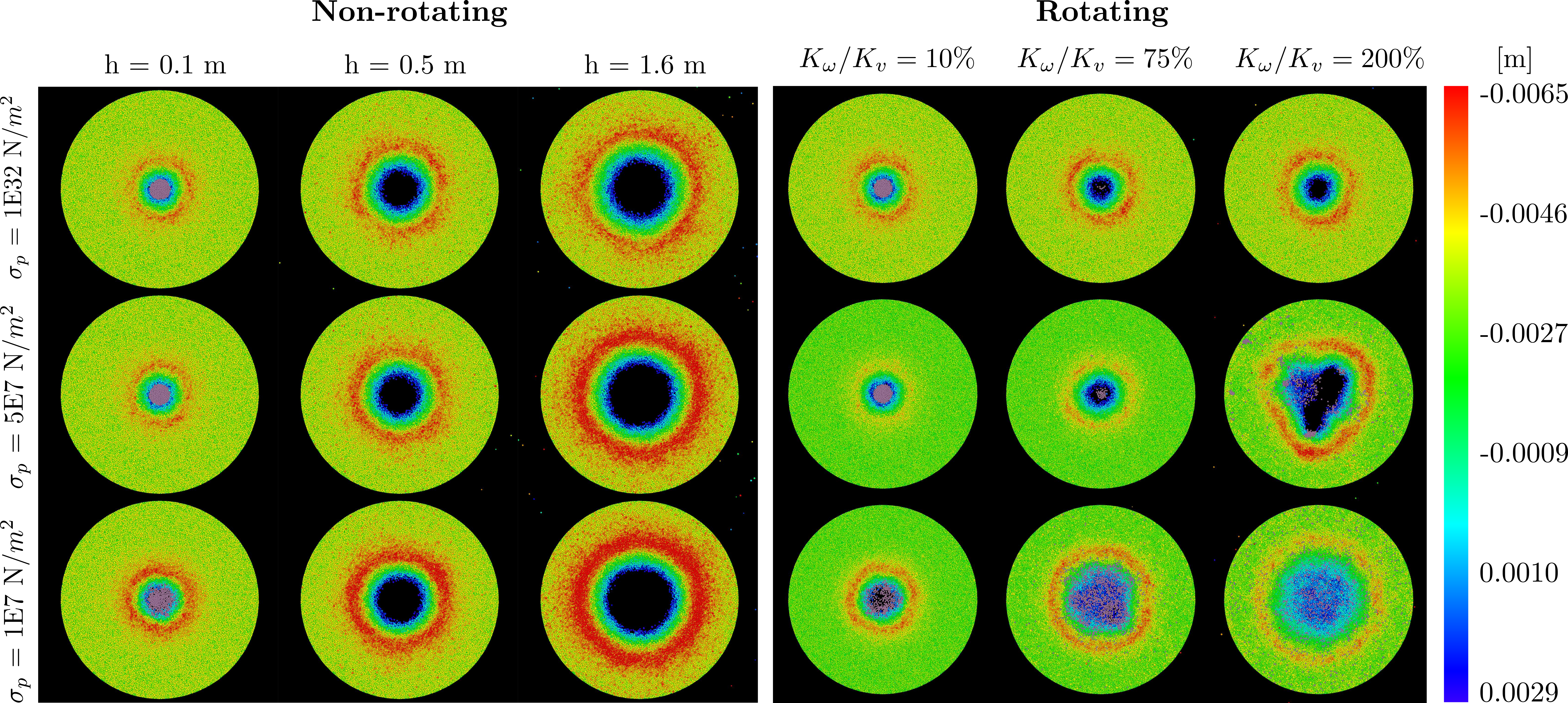}\\
	\end{center}
	\caption{Top view of final positions of grains, showing the final morphology of craters for non-rotating and rotating projectiles with different bonding stresses. For spinning projectiles, $h$ = 0.1 m. The colorbar on the right shows the elevation of each grain from the undisturbed surface (coordinate pointing downwards). The same figure in gray scale is available in the Supplemental Material \cite{Supplemental}}
	\label{fig:morphology}
\end{figure}

\begin{figure}[ht]
	\begin{center}
		\includegraphics[width=\linewidth]{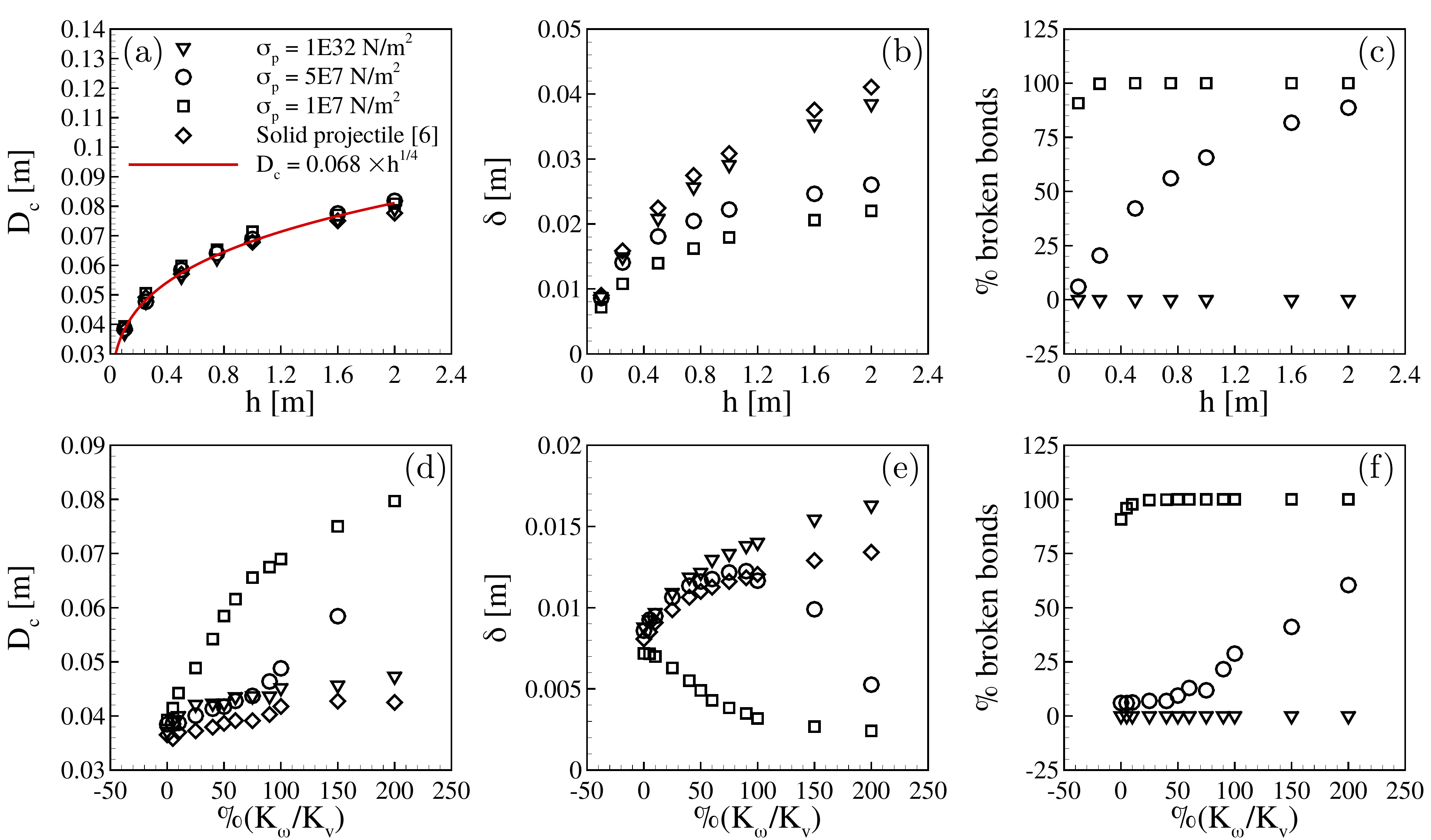}\\
	\end{center}
	\caption{(a) Crater diameter $D_c$, (b) penetration depth $\delta$, and (c) the percentage of broken bonds as functions of the initial height $h$ for a non-rotating projectile; panels (d), (e) and (f) show $D_c$, $\delta$ and the percentage of broken bonds as functions $K_{\omega}/K_v$ for spinning projectiles falling from $h$ = 0.1 m, respectively. The graphics are parameterized by the bonding stresses (shown in the key of panel (a)), and the results for the solid projectile reported in Carvalho et al. \cite{Carvalho} are shown for reference.}
	\label{fig:graphics}
\end{figure}

Figure \ref{fig:morphology} shows top view images of the final position of grains for non-rotating and rotating projectiles with different bonding stresses $\sigma_p$. The bonding stresses are listed on the left, the corresponding elevation (from the undisturbed surface) of each grain is shown on the right, and initial heights $h$ (non-rotating cases) and ratios of rotational to linear kinetic energies $K_{\omega}/K_v$ available at the impact (for spinning projectiles) are shown on the top. We used three different values of $\sigma_p$: $\sigma_p$ = 10$^{32}$ N/m$^2$, which is strong enough to ensure that the agglomerate behaves as a single solid (no breaking) for the range of energies simulated; $\sigma_p$ = 5 $\times$ 10$^7$ N/m$^2$, for which the aggregate collapses partially within the ranges of energy simulated; and $\sigma_p$ = 1 $\times$ 10$^7$ N/m$^2$, for which the projectile collapses completely for the highest energies simulated. For the non-rotating case, we observe that the crater diameter $D_c$ remains roughly constant and the height of the corona (rim) increases with the decrease in the bonding stresses, and, consequently, with the number of broken bonds (shown next in Fig. \ref{fig:graphics}(c)). In the rotating case, craters are shallower, wider, and with lower rims when compared to the non-rotating case. This is caused by the stronger spreading of grains when the projectile has an initial spin, which we inquire further in the following. In addition, we observe that large asymmetries can appear for rotating cases in which partial breaking occurs, such as when $\sigma_p$ = 5 $\times$ 10$^7$ N/m$^2$ and $K_{\omega}/K_v$ = 200\% (the partial breaking is confirmed in Fig. \ref{fig:graphics}(f)). The asymmetries come then from a small number of chunks spreading in the horizontal plane (when $\sigma_p$ = 5 $\times$ 10$^7$ N/m$^2$ and $K_{\omega}/K_v$ = 200\%, three large pieces were spread by centrifugal effect, see the Supplemental Material \cite{Supplemental} for snapshots of the final positions of grains originally in the projectile and a movie of the entire process).

\begin{figure}[ht]
	\begin{center}
		\includegraphics[width=\linewidth]{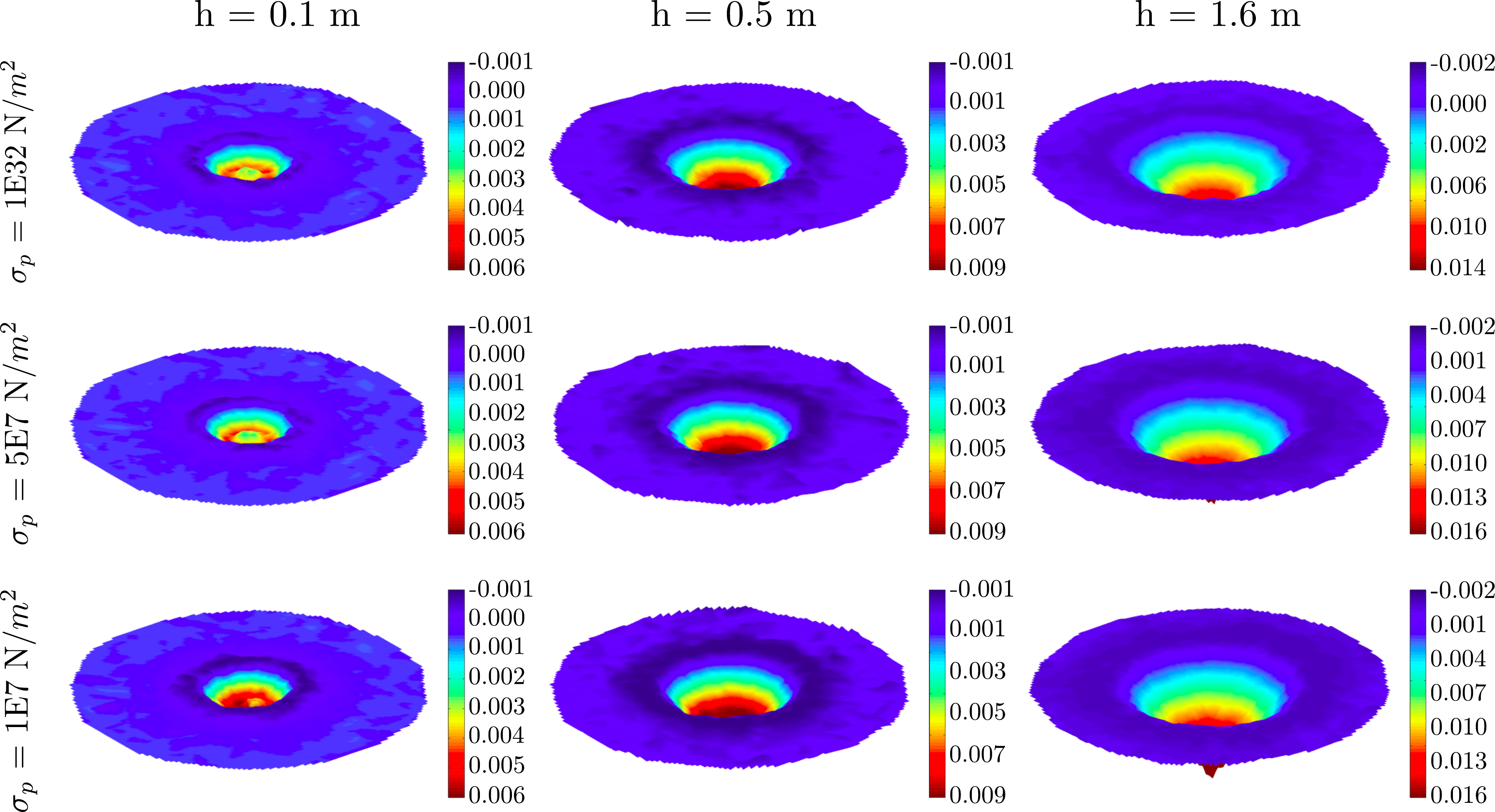}\\
	\end{center}
	\caption{Topography (elevation) of the final craters for non-rotating projectiles with different bonding stresses. The colorbar on the right of each panel shows the elevation from the undisturbed surface in m . The same figure in gray scale is available in the Supplemental Material \cite{Supplemental}}
	\label{fig:morphology2_nonrotating}
\end{figure}

\begin{figure}[ht]
	\begin{center}
		\includegraphics[width=\linewidth]{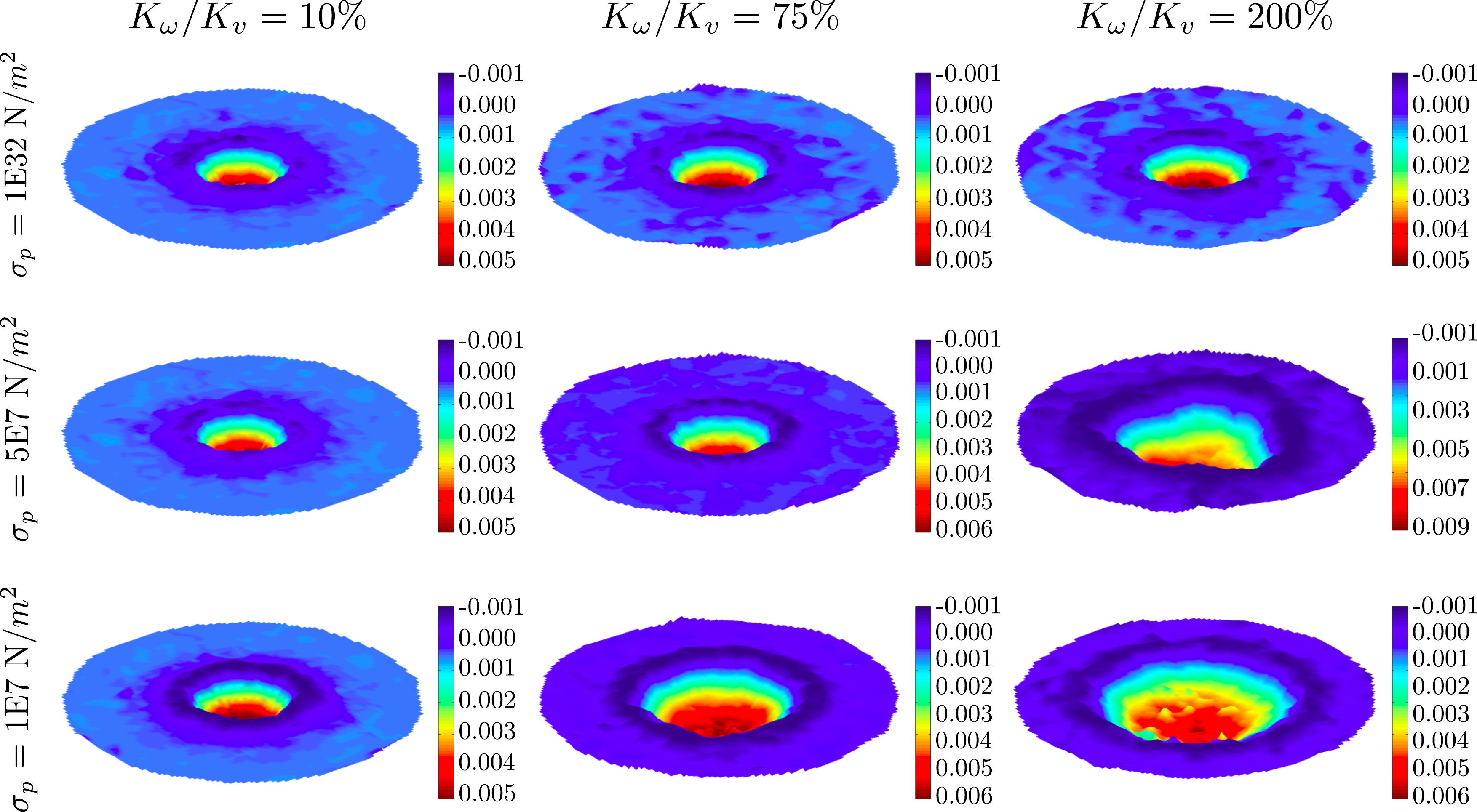}\\
	\end{center}
	\caption{Topography (elevation) of the final craters for rotating projectiles with different bonding stresses. The colorbar on the right of each panel shows the elevation from the undisturbed surface in m, and $h$ = 0.1 m for all panels. The same figure in gray scale is available in the Supplemental Material \cite{Supplemental}}
	\label{fig:morphology2_rotating}
\end{figure}

\begin{figure}[ht]
	\begin{center}
		\includegraphics[width=\linewidth]{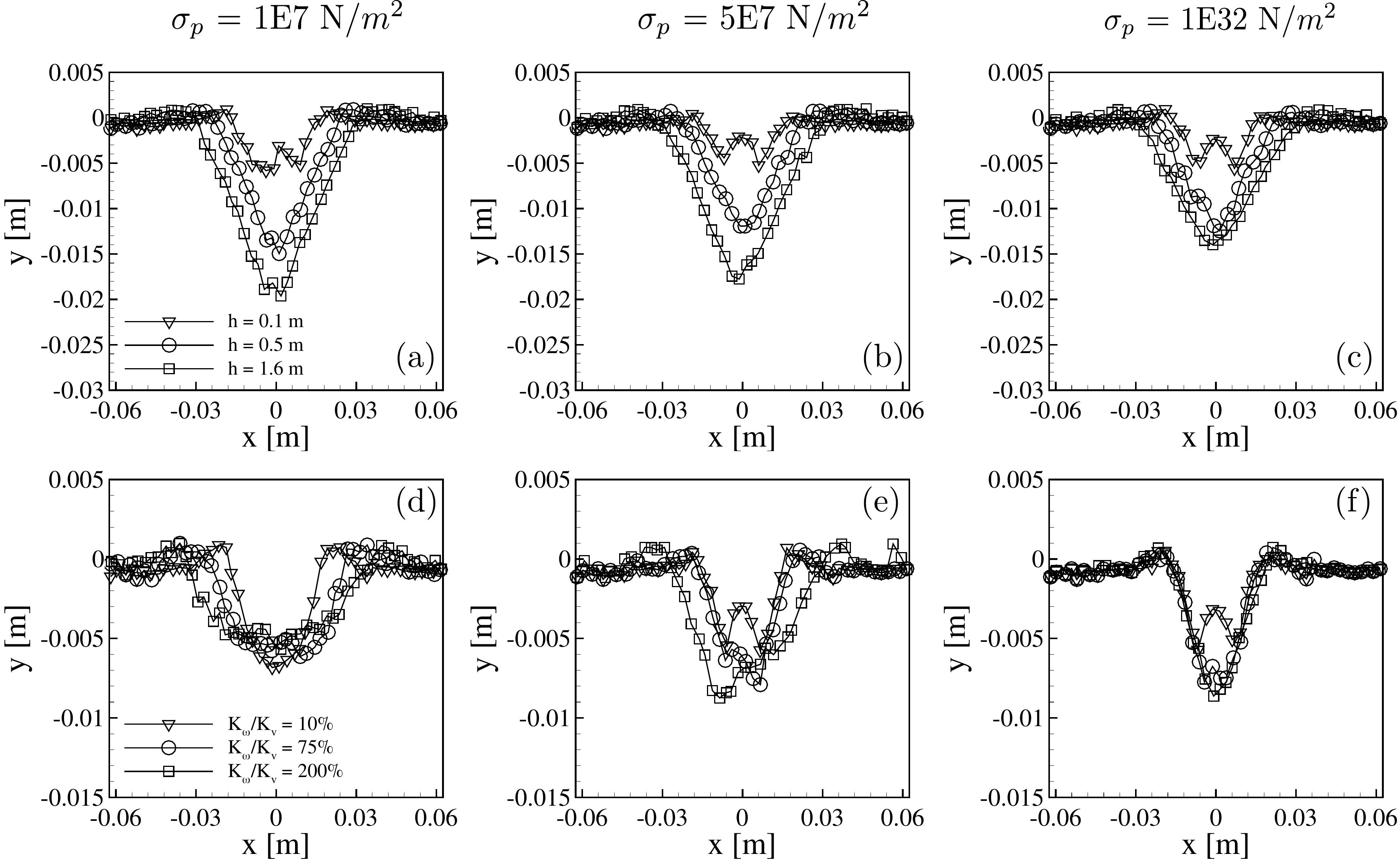}\\
	\end{center}
	\caption{Profiles of the elevations of final craters for both non-rotating and rotating projectiles, with different bonding stresses. The heights and rotational energies are shown in the figure key, and $h$ = 0.1 m for non-rotating projectiles. All profiles were plot in a vertical plane of symmetry (therefore, include the crater center). These profiles include the projectile's grains (see the Supplemental Material \cite{Supplemental} for profiles excluding the projectile's grains).}
	\label{fig:crater_profile}
\end{figure}

Most of the aforementioned comments can be observed in Figs. \ref{fig:morphology2_nonrotating} and \ref{fig:morphology2_rotating}, which show the topography (elevation) of the final craters for non-rotating and rotating projectiles, respectively, for the same variations of the bonding stress and available energy of Fig. \ref{fig:morphology}. Although variations of $D_c$ are easier observed in Fig. \ref{fig:morphology}, Figs. \ref{fig:morphology2_nonrotating} and \ref{fig:morphology2_rotating} allow for easier and direct observations of the crater depth and the formation of small peaks (we note that the scales of figures are not the same). We notice that the crater shape becomes flatter, with peaks around the rim and in the center of craters as the bonding stresses decrease and the initial spin increases (although peaks can also appear in low-energy cases without fragmentation). Some of these observations are corroborated by Fig. \ref{fig:crater_profile}, which shows profiles of the elevations of final craters for both non-rotating and rotating projectiles, with different bonding stresses. Profiles corresponding to different heights are superimposed for non-rotating cases and to different rotational energies for rotating cases, allowing direct comparisons. We observe that craters have higher diameters and lower depths when projectiles have large rotational energies and low bonding stresses, and that some oscillations appear in the region near the corona (corresponding to peripheral peaks). We can observe a central peak in low-energy non-fragmenting cases, but they correspond to the projectile itself (which was not completely buried, see the Supplemental Material \cite{Supplemental} for profiles excluding the projectile's grains). Therefore, the final topographies indicate that the formation of central and peripheral peaks are due to the stronger spreading of grains when the projectile has higher rotational energies. In addition, the central peak can also be formed by a partially penetrating projectile when the available energy is relatively low.

In order to inquire further into the crater shape and the level of fracture of the projectile, we plot in Fig. \ref{fig:graphics} the crater diameter $D_c$, the penetration depth $\delta$, and the percentage of broken bonds as functions of the initial height $h$ or the ratio of rotational to linear kinetic energies $K_{\omega}/K_v$ for, respectively, non-rotating and rotating projectiles. The crater diameter $D_c$ was determined as the diameter of a circle fitted over the corona, and corresponds to an equivalent diameter in the case of asymmetric craters. Whenever the projectile collapsed, we computed $\delta$ based on the center of mass of the projectile's grains. For the non-rotating case, we observe that $D_c$ (Fig. \ref{fig:graphics}(a)) is roughly independent of $\sigma_p$ (for the levels of energy investigated in this paper), varying as $D_c \sim h^{1/4}$, in agreement with Pacheco-V\'azquez and Ruiz-Su\'arez \cite{Pacheco2}, although they measured the packing fraction of agglomerates instead of $\sigma_p$ (to which we have access in our simulations). However, Pacheco-V\'azquez and Ruiz-Su\'arez \cite{Pacheco2} identified a discontinuity in $D_c$ as a result of fragmentation, which depended on the projectile packing fraction. We did not observe the discontinuity, perhaps because our projectiles were lighter than those in Ref. \cite{Pacheco2} (13.9 g in our simulations, against 33.0--45.5g in their experiments). The penetration depth $\delta$ (Fig. \ref{fig:graphics}(b)), on the other hand, depends on $\sigma_p$, varying with the percentage of broken bonds (Fig. \ref{fig:graphics}(c)). In addition, the rate of change of $\delta$ with $h$ decreases as $h$ increases, and it is possible that a plateau is reached for values of $h$ higher than those simulated in this work. This would be in agreement with the results of Ref. \cite{Pacheco2}, but remains to be investigated further. For $\delta$, Pacheco-V\'azquez and Ruiz-Su\'arez \cite{Pacheco2} also found a discontinuity resulting from fragmentation, which our simulations did not show. As stated for $D_c$, the absence of discontinuity is due, perhaps, to the lighter weight of our projectiles. For the rotating case, the situation is different: $D_c$ varies strongly with $\sigma_p$ (Fig. \ref{fig:graphics}(d)), and variations for $\delta$ are even stronger (Fig. \ref{fig:graphics}(e)). Figure \ref{fig:graphics}(d) shows that $D_c$ increases up to approximately 2 times when $\sigma_p$ varies from the largest (non-breaking) to the lowest (breaking) values (for $K_{\omega}/K_v$ varying between 0 and 200\%), and for moderate stresses ($\sigma_p$ = 5 $\times$ 10$^7$ N/m$^2$) we notice that partial breaking makes $D_c$ to deviate from the curve for the non-breaking case toward to that for the breaking case (which occurs for $K_{\omega}/K_v$ around 100\% in Fig. \ref{fig:graphics}(d)). The inverse behavior occurs for $\delta$: it decreases by one order of magnitude when $\sigma_p$ varies from the largest to the lowest value, with also partial breaking ($\sigma_p$ = 5 $\times$ 10$^7$ N/m$^2$) leading to the breaking case as $K_{\omega}/K_v$ increases. Finally, Fig. \ref{fig:graphics}(f) shows that, indeed, the percentage of broken bonds is 0\% for the largest $\sigma_p$, and 100\% for the lowest $\sigma_p$ when $K_{\omega}/K_v$ $\geq$ 30\%, while that for moderate $\sigma_p$ evolves toward 100\% for increasing $K_{\omega}/K_v$. At the same time, values of $\delta$ for $\sigma_p$ = 5 $\times$ 10$^7$ N/m$^2$ evolve toward those for $\sigma_p$ = 1 $\times$ 10$^7$ N/m$^2$ (Fig.\ref{fig:graphics}(e)). This implies that the penetration depth of rotating projectiles varies with their angular velocity and degree of collapse (number of detached particles), but not necessarily with the bonding stresses, indicating that under high spinning velocities the excess of breaking energy contributes only for the larger spreading in the horizontal plane and the formation of peaks.

\begin{figure}[ht]
	\begin{center}
		\includegraphics[width=\linewidth]{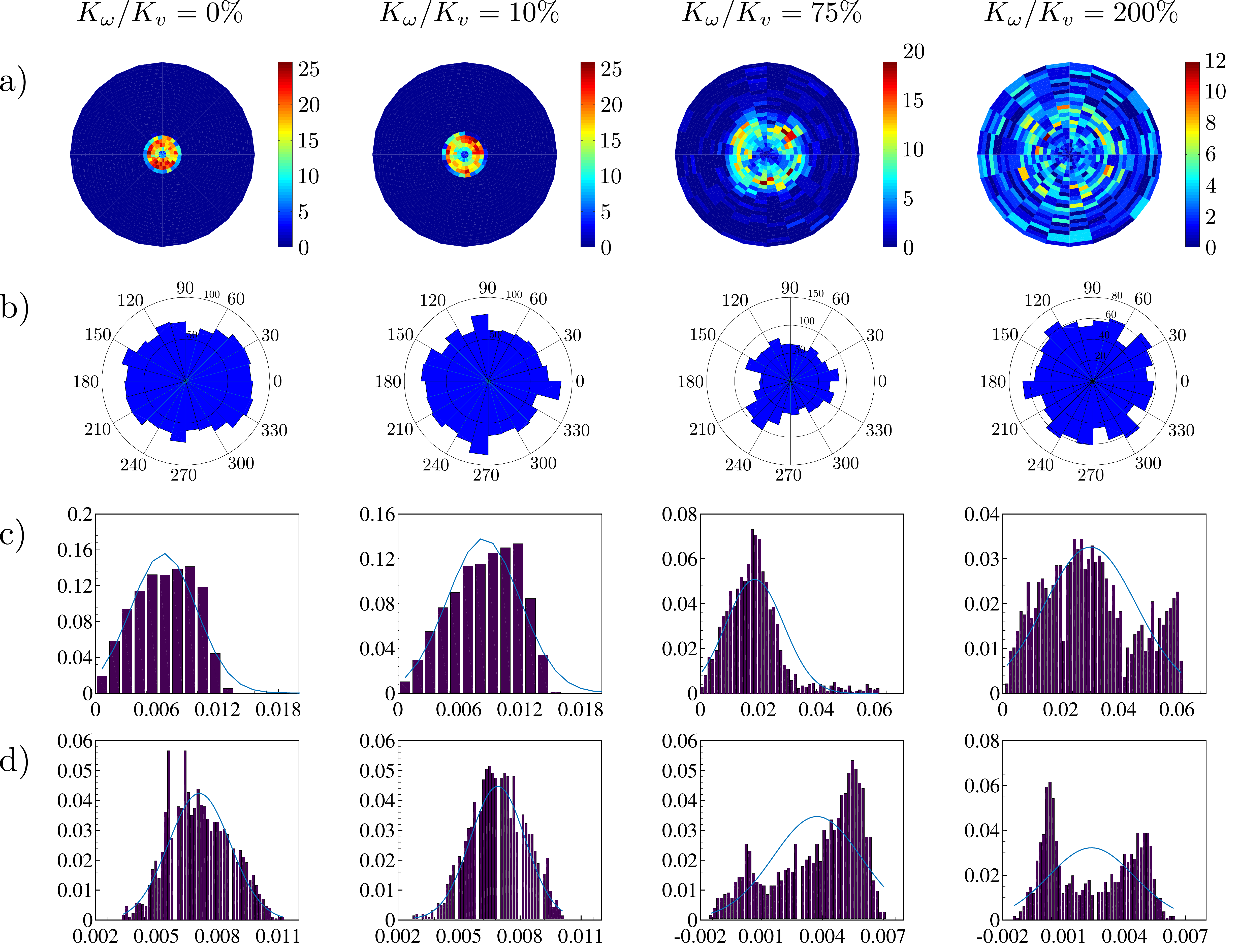}\\
	\end{center}
	\caption{Final positions the projectile's grains after the impact has taken place, for $\sigma_p$ = 1 $\times$ 10$^7$ N/m$^2$ and different values of $K_{\omega}/K_v$. From top to bottom: (a) Frequencies of occurrence of the projectile's grains in the $r$--$\theta$ plane (radius--angle plane, independent of the depth); (b) frequencies of occurrence of final positions in terms of the angle (all depths); (c) frequencies of occurrence of final positions in terms of radius (all depths); (d) frequencies of occurrence of final positions in the $y$ coordinate (depths for all angles and radii).}
	\label{fig:morphology3}
\end{figure}

After the impact has occurred, the projectile collapses if the bonding stresses are not strong enough to maintain the agglomerate integrity. In these cases, besides changing considerably the crater shape, the once agglomerated material is spread on or within the ground, over distances that depend on the initial height, bonding stresses and initial spin of the projectile. Understanding how this process occurs can help us, for example, to interpret whether materials found today under the ground have their origin on the ancient impact of asteroids, and how they are distributed, with important applications in geophysics and mining. Therefore, we inquire now into the dispersion of the projectile's grains.

Figure \ref{fig:morphology3} shows the final positions of grains initially forming the projectile, for $\sigma_p$ = 1 $\times$ 10$^7$ N/m$^2$ and different $K_{\omega}/K_v$. The first row (Fig. \ref{fig:morphology3}(a)) shows the frequencies of occurrence of the projectile's grains in the $r$--$\theta$ plane (radius--angle plane, independent of the depth), corresponding then to top views of the distributions of the projectile's grains (in the polar plane). We clearly observe that the projectile material reaches distances farther from the collision point as the rotational energy increases. In order to have more quantitative measurements, the second (Fig. \ref{fig:morphology3}(b)) and third (Fig. \ref{fig:morphology3}(c)) rows show the frequencies of occurrence of final positions in terms of the angle $\theta$ and radius $r$, for all depths, where the angles are given in degrees and the radius in m (see Fig. \ref{fig:craters}(c) for $\theta$ and $r$). We observe that in this weak-bond case the projectile's grains spread in a roughly symmetrical way along the angles, and distances reached in the radial direction increase with $K_{\omega}/K_v$: the most probable values of the radius increase from approximately 0.007 m when $K_{\omega}/K_v$ = 0 to 0.03 m (one order of magnitude greater) when $K_{\omega}/K_v$ = 200\%. Finally, the last row (Fig. \ref{fig:morphology3}(d)) shows the frequencies of occurrence of final positions in terms of depth (all angles and radii). Interestingly, we observe that the most probable value decreases with $K_{\omega}/K_v$, so that in average the projectile's grains tend to remain closer to the surface for higher spinning velocities, different from the behavior of solid projectiles (which reach deeper depths for increasing $K_{\omega}/K_v$, Ref. \cite{Carvalho}). However, the depth distribution widens, so that the projectile's grains populate depths that spans over larger values, including negative ones corresponding to peaks or the corona. Snapshots of the final positions of grains originally in the projectile are available in the Supplemental Material \cite{Supplemental}. 

We note that we did not investigate the effect of initial packing fractions on the dynamics of cratering in this paper (it was the object of Carvalho et al. \cite{Carvalho}). However, we measured how the bed packing fraction far from the collision point varies with the linear and rotational energies, for the different bonding stresses used. For that, we selected a 20-mm-height cylindrical region occupying the bottom of the cylindrical container (corresponding to 26\% of the container) and measured the average packing fraction before and after the impact. For rotating projectiles, we found no change at all in the packing fraction, while negligible variations (increasing with $h$) were measured for non-rotating projectiles. The maximum variations were of 0.34\%, 0.30\% and 0.20\% for $\sigma_p$ = 10$^7$, 5 $\times$ 10$^7$ and 10$^{32}$, respectively, and $h$ = 2 m. We also note that, under some conditions, the dynamics of both cratering and projectile fragmentation change with the stiffness of grains and bonds. Due to the presence of bonds, the effect of stiffness is rather complex and needs to be investigated further.

\section{CONCLUSIONS}
\label{sec:conclusions}

In this paper, we investigated numerically how the projectile spin and cohesion lead to different crater shapes, and how the projectile's materials spread over and below the ground. We found that, as the bonding stresses decrease and the initial spin increases: (i) the projectile's grains spread radially farther from the collision point; (ii) the projectile's grains remain in average closer to the surface (lower penetration depths), but spread horizontally over longer distances, with some grains buried deep in the bed while others are above the surface populating peaks or the corona; (iii) as a consequence, the crater shape becomes flatter, with peaks around the rim and in the center of craters. In addition, we found that the penetration depth of rotating projectiles varies with their angular velocity and degree of collapse (number of detached particles), but not necessarily with the bonding stresses themselves, indicating that under high spinning velocities the excess of breaking energy contributes only for the larger spreading in the horizontal plane and formation of peaks. Our results represent a significant step for understanding how cratering occurs, helping us, for example, to interpret whether materials found today under the ground have their origin on the ancient impact of asteroids, and how they are distributed, with important applications in geophysics and engineering.

\section{\label{sec:Ack} ACKNOWLEDGMENTS}

The authors are grateful to the S\~ao Paulo Research Foundation -- FAPESP (Grants No. 2018/14981-7, No. 2019/20888-2 and No. 2020/04151-7) and Conselho Nacional de Desenvolvimento Cient\'ifico e Tecnol\'ogico -- CNPq (Grant No. 405512/2022-8) for the financial support provided.

\bibliography{references}

\begin{thebibliography}{37}%
\makeatletter
\providecommand \@ifxundefined [1]{%
 \@ifx{#1\undefined}
}%
\providecommand \@ifnum [1]{%
 \ifnum #1\expandafter \@firstoftwo
 \else \expandafter \@secondoftwo
 \fi
}%
\providecommand \@ifx [1]{%
 \ifx #1\expandafter \@firstoftwo
 \else \expandafter \@secondoftwo
 \fi
}%
\providecommand \natexlab [1]{#1}%
\providecommand \enquote  [1]{``#1''}%
\providecommand \bibnamefont  [1]{#1}%
\providecommand \bibfnamefont [1]{#1}%
\providecommand \citenamefont [1]{#1}%
\providecommand \href@noop [0]{\@secondoftwo}%
\providecommand \href [0]{\begingroup \@sanitize@url \@href}%
\providecommand \@href[1]{\@@startlink{#1}\@@href}%
\providecommand \@@href[1]{\endgroup#1\@@endlink}%
\providecommand \@sanitize@url [0]{\catcode `\\12\catcode `\$12\catcode
  `\&12\catcode `\#12\catcode `\^12\catcode `\_12\catcode `\%12\relax}%
\providecommand \@@startlink[1]{}%
\providecommand \@@endlink[0]{}%
\providecommand \url  [0]{\begingroup\@sanitize@url \@url }%
\providecommand \@url [1]{\endgroup\@href {#1}{\urlprefix }}%
\providecommand \urlprefix  [0]{URL }%
\providecommand \Eprint [0]{\href }%
\providecommand \doibase [0]{https://doi.org/}%
\providecommand \selectlanguage [0]{\@gobble}%
\providecommand \bibinfo  [0]{\@secondoftwo}%
\providecommand \bibfield  [0]{\@secondoftwo}%
\providecommand \translation [1]{[#1]}%
\providecommand \BibitemOpen [0]{}%
\providecommand \bibitemStop [0]{}%
\providecommand \bibitemNoStop [0]{.\EOS\space}%
\providecommand \EOS [0]{\spacefactor3000\relax}%
\providecommand \BibitemShut  [1]{\csname bibitem#1\endcsname}%
\let\auto@bib@innerbib\@empty
\bibitem [{\citenamefont {Melosh}\ and\ \citenamefont {Ivanov}(1999)}]{Melosh}%
  \BibitemOpen
  \bibfield  {author} {\bibinfo {author} {\bibfnamefont {H.~J.}\ \bibnamefont
  {Melosh}}\ and\ \bibinfo {author} {\bibfnamefont {B.~A.}\ \bibnamefont
  {Ivanov}},\ }\bibfield  {title} {\bibinfo {title} {Impact crater collapse},\
  }\href@noop {} {\bibfield  {journal} {\bibinfo  {journal} {Ann. Rev. Earth
  Pl. Sc.}\ }\textbf {\bibinfo {volume} {27}},\ \bibinfo {pages} {385}
  (\bibinfo {year} {1999})}\BibitemShut {NoStop}%
\bibitem [{\citenamefont {Barlow}\ \emph {et~al.}(2017)\citenamefont {Barlow},
  \citenamefont {Ferguson}, \citenamefont {Horstman},\ and\ \citenamefont
  {Maine}}]{Barlow}%
  \BibitemOpen
  \bibfield  {author} {\bibinfo {author} {\bibfnamefont {N.~G.}\ \bibnamefont
  {Barlow}}, \bibinfo {author} {\bibfnamefont {S.~N.}\ \bibnamefont
  {Ferguson}}, \bibinfo {author} {\bibfnamefont {R.~M.}\ \bibnamefont
  {Horstman}},\ and\ \bibinfo {author} {\bibfnamefont {A.}~\bibnamefont
  {Maine}},\ }\bibfield  {title} {\bibinfo {title} {Comparison of central pit
  craters on mars, mercury, ganymede, and the saturnian satellites},\
  }\href@noop {} {\bibfield  {journal} {\bibinfo  {journal} {Meteorit. Planet.
  Sci.}\ }\textbf {\bibinfo {volume} {52}},\ \bibinfo {pages} {1371} (\bibinfo
  {year} {2017})}\BibitemShut {NoStop}%
\bibitem [{\citenamefont {Arvidson}(1974)}]{Arvidson}%
  \BibitemOpen
  \bibfield  {author} {\bibinfo {author} {\bibfnamefont {R.~E.}\ \bibnamefont
  {Arvidson}},\ }\bibfield  {title} {\bibinfo {title} {Morphologic
  classification of martian craters and some implications},\ }\href@noop {}
  {\bibfield  {journal} {\bibinfo  {journal} {Icarus}\ }\textbf {\bibinfo
  {volume} {22}},\ \bibinfo {pages} {264} (\bibinfo {year} {1974})}\BibitemShut
  {NoStop}%
\bibitem [{\citenamefont {Pacheco-V\'azquez}\ and\ \citenamefont
  {Ruiz-Su\'arez}(2011)}]{Pacheco2}%
  \BibitemOpen
  \bibfield  {author} {\bibinfo {author} {\bibfnamefont {F.}~\bibnamefont
  {Pacheco-V\'azquez}}\ and\ \bibinfo {author} {\bibfnamefont {J.~C.}\
  \bibnamefont {Ruiz-Su\'arez}},\ }\bibfield  {title} {\bibinfo {title} {Impact
  craters in granular media: Grains against grains},\ }\href
  {https://doi.org/10.1103/PhysRevLett.107.218001} {\bibfield  {journal}
  {\bibinfo  {journal} {Phys. Rev. Lett.}\ }\textbf {\bibinfo {volume} {107}},\
  \bibinfo {pages} {218001} (\bibinfo {year} {2011})}\BibitemShut {NoStop}%
\bibitem [{\citenamefont {Seguin}\ \emph {et~al.}(2008)\citenamefont {Seguin},
  \citenamefont {Bertho},\ and\ \citenamefont {Gondret}}]{Seguin2}%
  \BibitemOpen
  \bibfield  {author} {\bibinfo {author} {\bibfnamefont {A.}~\bibnamefont
  {Seguin}}, \bibinfo {author} {\bibfnamefont {Y.}~\bibnamefont {Bertho}},\
  and\ \bibinfo {author} {\bibfnamefont {P.}~\bibnamefont {Gondret}},\
  }\bibfield  {title} {\bibinfo {title} {Influence of confinement on granular
  penetration by impact},\ }\href {https://doi.org/10.1103/PhysRevE.78.010301}
  {\bibfield  {journal} {\bibinfo  {journal} {Phys. Rev. E}\ }\textbf {\bibinfo
  {volume} {78}},\ \bibinfo {pages} {010301} (\bibinfo {year}
  {2008})}\BibitemShut {NoStop}%
\bibitem [{\citenamefont {Carvalho}\ \emph {et~al.}(2023)\citenamefont
  {Carvalho}, \citenamefont {Lima},\ and\ \citenamefont {Franklin}}]{Carvalho}%
  \BibitemOpen
  \bibfield  {author} {\bibinfo {author} {\bibfnamefont {D.~D.}\ \bibnamefont
  {Carvalho}}, \bibinfo {author} {\bibfnamefont {N.~C.}\ \bibnamefont {Lima}},\
  and\ \bibinfo {author} {\bibfnamefont {E.~M.}\ \bibnamefont {Franklin}},\
  }\bibfield  {title} {\bibinfo {title} {Roles of packing fraction, microscopic
  friction, and projectile spin in cratering by impact},\ }\href
  {https://doi.org/10.1103/PhysRevE.107.044901} {\bibfield  {journal} {\bibinfo
   {journal} {Phys. Rev. E}\ }\textbf {\bibinfo {volume} {107}},\ \bibinfo
  {pages} {044901} (\bibinfo {year} {2023})}\BibitemShut {NoStop}%
\bibitem [{\citenamefont {Holsapple}(1993)}]{Holsapple}%
  \BibitemOpen
  \bibfield  {author} {\bibinfo {author} {\bibfnamefont {K.~A.}\ \bibnamefont
  {Holsapple}},\ }\bibfield  {title} {\bibinfo {title} {The scaling of impact
  processes in planetary sciences},\ }\href@noop {} {\bibfield  {journal}
  {\bibinfo  {journal} {Ann. Rev. Earth Pl. Sc.}\ }\textbf {\bibinfo {volume}
  {21}},\ \bibinfo {pages} {333} (\bibinfo {year} {1993})}\BibitemShut
  {NoStop}%
\bibitem [{\citenamefont {Uehara}\ \emph
  {et~al.}(2003{\natexlab{a}})\citenamefont {Uehara}, \citenamefont {Ambroso},
  \citenamefont {Ojha},\ and\ \citenamefont {Durian}}]{Uehara}%
  \BibitemOpen
  \bibfield  {author} {\bibinfo {author} {\bibfnamefont {J.~S.}\ \bibnamefont
  {Uehara}}, \bibinfo {author} {\bibfnamefont {M.~A.}\ \bibnamefont {Ambroso}},
  \bibinfo {author} {\bibfnamefont {R.~P.}\ \bibnamefont {Ojha}},\ and\
  \bibinfo {author} {\bibfnamefont {D.~J.}\ \bibnamefont {Durian}},\ }\bibfield
   {title} {\bibinfo {title} {Low-speed impact craters in loose granular
  media},\ }\href {https://doi.org/10.1103/PhysRevLett.90.194301} {\bibfield
  {journal} {\bibinfo  {journal} {Phys. Rev. Lett.}\ }\textbf {\bibinfo
  {volume} {90}},\ \bibinfo {pages} {194301} (\bibinfo {year}
  {2003}{\natexlab{a}})}\BibitemShut {NoStop}%
\bibitem [{\citenamefont {Walsh}\ \emph {et~al.}(2003)\citenamefont {Walsh},
  \citenamefont {Holloway}, \citenamefont {Habdas},\ and\ \citenamefont
  {de~Bruyn}}]{Walsh}%
  \BibitemOpen
  \bibfield  {author} {\bibinfo {author} {\bibfnamefont {A.~M.}\ \bibnamefont
  {Walsh}}, \bibinfo {author} {\bibfnamefont {K.~E.}\ \bibnamefont {Holloway}},
  \bibinfo {author} {\bibfnamefont {P.}~\bibnamefont {Habdas}},\ and\ \bibinfo
  {author} {\bibfnamefont {J.~R.}\ \bibnamefont {de~Bruyn}},\ }\bibfield
  {title} {\bibinfo {title} {Morphology and scaling of impact craters in
  granular media},\ }\href {https://doi.org/10.1103/PhysRevLett.91.104301}
  {\bibfield  {journal} {\bibinfo  {journal} {Phys. Rev. Lett.}\ }\textbf
  {\bibinfo {volume} {91}},\ \bibinfo {pages} {104301} (\bibinfo {year}
  {2003})}\BibitemShut {NoStop}%
\bibitem [{\citenamefont {Ciamarra}\ \emph {et~al.}(2004)\citenamefont
  {Ciamarra}, \citenamefont {Lara}, \citenamefont {Lee}, \citenamefont
  {Goldman}, \citenamefont {Vishik},\ and\ \citenamefont {Swinney}}]{Ciamarra}%
  \BibitemOpen
  \bibfield  {author} {\bibinfo {author} {\bibfnamefont {M.~P.}\ \bibnamefont
  {Ciamarra}}, \bibinfo {author} {\bibfnamefont {A.~H.}\ \bibnamefont {Lara}},
  \bibinfo {author} {\bibfnamefont {A.~T.}\ \bibnamefont {Lee}}, \bibinfo
  {author} {\bibfnamefont {D.~I.}\ \bibnamefont {Goldman}}, \bibinfo {author}
  {\bibfnamefont {I.}~\bibnamefont {Vishik}},\ and\ \bibinfo {author}
  {\bibfnamefont {H.~L.}\ \bibnamefont {Swinney}},\ }\bibfield  {title}
  {\bibinfo {title} {Dynamics of drag and force distributions for projectile
  impact in a granular medium},\ }\href
  {https://doi.org/10.1103/PhysRevLett.92.194301} {\bibfield  {journal}
  {\bibinfo  {journal} {Phys. Rev. Lett.}\ }\textbf {\bibinfo {volume} {92}},\
  \bibinfo {pages} {194301} (\bibinfo {year} {2004})}\BibitemShut {NoStop}%
\bibitem [{\citenamefont {Katsuragi}\ and\ \citenamefont
  {Durian}(2007)}]{Katsuragi}%
  \BibitemOpen
  \bibfield  {author} {\bibinfo {author} {\bibfnamefont {H.}~\bibnamefont
  {Katsuragi}}\ and\ \bibinfo {author} {\bibfnamefont {D.}~\bibnamefont
  {Durian}},\ }\bibfield  {title} {\bibinfo {title} {Unified force law for
  granular impact cratering},\ }\href@noop {} {\bibfield  {journal} {\bibinfo
  {journal} {Nature Phys.}\ }\textbf {\bibinfo {volume} {3}},\ \bibinfo {pages}
  {420} (\bibinfo {year} {2007})}\BibitemShut {NoStop}%
\bibitem [{\citenamefont {de~Vet}\ and\ \citenamefont
  {de~Bruyn}(2007)}]{deVet}%
  \BibitemOpen
  \bibfield  {author} {\bibinfo {author} {\bibfnamefont {S.~J.}\ \bibnamefont
  {de~Vet}}\ and\ \bibinfo {author} {\bibfnamefont {J.~R.}\ \bibnamefont
  {de~Bruyn}},\ }\bibfield  {title} {\bibinfo {title} {Shape of impact craters
  in granular media},\ }\href {https://doi.org/10.1103/PhysRevE.76.041306}
  {\bibfield  {journal} {\bibinfo  {journal} {Phys. Rev. E}\ }\textbf {\bibinfo
  {volume} {76}},\ \bibinfo {pages} {041306} (\bibinfo {year}
  {2007})}\BibitemShut {NoStop}%
\bibitem [{\citenamefont {Goldman}\ and\ \citenamefont
  {Umbanhowar}(2008)}]{Goldman}%
  \BibitemOpen
  \bibfield  {author} {\bibinfo {author} {\bibfnamefont {D.~I.}\ \bibnamefont
  {Goldman}}\ and\ \bibinfo {author} {\bibfnamefont {P.}~\bibnamefont
  {Umbanhowar}},\ }\bibfield  {title} {\bibinfo {title} {Scaling and dynamics
  of sphere and disk impact into granular media},\ }\href
  {https://doi.org/10.1103/PhysRevE.77.021308} {\bibfield  {journal} {\bibinfo
  {journal} {Phys. Rev. E}\ }\textbf {\bibinfo {volume} {77}},\ \bibinfo
  {pages} {021308} (\bibinfo {year} {2008})}\BibitemShut {NoStop}%
\bibitem [{\citenamefont {Seguin}\ \emph {et~al.}(2009)\citenamefont {Seguin},
  \citenamefont {Bertho}, \citenamefont {Gondret},\ and\ \citenamefont
  {Crassous}}]{Seguin}%
  \BibitemOpen
  \bibfield  {author} {\bibinfo {author} {\bibfnamefont {A.}~\bibnamefont
  {Seguin}}, \bibinfo {author} {\bibfnamefont {Y.}~\bibnamefont {Bertho}},
  \bibinfo {author} {\bibfnamefont {P.}~\bibnamefont {Gondret}},\ and\ \bibinfo
  {author} {\bibfnamefont {J.}~\bibnamefont {Crassous}},\ }\bibfield  {title}
  {\bibinfo {title} {Sphere penetration by impact in a granular medium: A
  collisional process},\ }\href@noop {} {\bibfield  {journal} {\bibinfo
  {journal} {{EPL} (Europhysics Letters)}\ }\textbf {\bibinfo {volume} {88}},\
  \bibinfo {pages} {44002} (\bibinfo {year} {2009})}\BibitemShut {NoStop}%
\bibitem [{\citenamefont {Umbanhowar}\ and\ \citenamefont
  {Goldman}(2010)}]{Umbanhowar}%
  \BibitemOpen
  \bibfield  {author} {\bibinfo {author} {\bibfnamefont {P.}~\bibnamefont
  {Umbanhowar}}\ and\ \bibinfo {author} {\bibfnamefont {D.~I.}\ \bibnamefont
  {Goldman}},\ }\bibfield  {title} {\bibinfo {title} {Granular impact and the
  critical packing state},\ }\href {https://doi.org/10.1103/PhysRevE.82.010301}
  {\bibfield  {journal} {\bibinfo  {journal} {Phys. Rev. E}\ }\textbf {\bibinfo
  {volume} {82}},\ \bibinfo {pages} {010301} (\bibinfo {year}
  {2010})}\BibitemShut {NoStop}%
\bibitem [{\citenamefont {Katsuragi}\ and\ \citenamefont
  {Durian}(2013)}]{Katsuragi2}%
  \BibitemOpen
  \bibfield  {author} {\bibinfo {author} {\bibfnamefont {H.}~\bibnamefont
  {Katsuragi}}\ and\ \bibinfo {author} {\bibfnamefont {D.~J.}\ \bibnamefont
  {Durian}},\ }\bibfield  {title} {\bibinfo {title} {Drag force scaling for
  penetration into granular media},\ }\href
  {https://doi.org/10.1103/PhysRevE.87.052208} {\bibfield  {journal} {\bibinfo
  {journal} {Phys. Rev. E}\ }\textbf {\bibinfo {volume} {87}},\ \bibinfo
  {pages} {052208} (\bibinfo {year} {2013})}\BibitemShut {NoStop}%
\bibitem [{\citenamefont {Ruiz-Su{\'{a}}rez}(2013)}]{Suarez}%
  \BibitemOpen
  \bibfield  {author} {\bibinfo {author} {\bibfnamefont {J.~C.}\ \bibnamefont
  {Ruiz-Su{\'{a}}rez}},\ }\bibfield  {title} {\bibinfo {title} {Penetration of
  projectiles into granular targets},\ }\href@noop {} {\bibfield  {journal}
  {\bibinfo  {journal} {Rep. Prog. Phys.}\ }\textbf {\bibinfo {volume} {76}},\
  \bibinfo {pages} {066601} (\bibinfo {year} {2013})}\BibitemShut {NoStop}%
\bibitem [{\citenamefont {Uehara}\ \emph
  {et~al.}(2003{\natexlab{b}})\citenamefont {Uehara}, \citenamefont {Ambroso},
  \citenamefont {Ojha},\ and\ \citenamefont {Durian}}]{Uehara2}%
  \BibitemOpen
  \bibfield  {author} {\bibinfo {author} {\bibfnamefont {J.~S.}\ \bibnamefont
  {Uehara}}, \bibinfo {author} {\bibfnamefont {M.~A.}\ \bibnamefont {Ambroso}},
  \bibinfo {author} {\bibfnamefont {R.~P.}\ \bibnamefont {Ojha}},\ and\
  \bibinfo {author} {\bibfnamefont {D.~J.}\ \bibnamefont {Durian}},\ }\bibfield
   {title} {\bibinfo {title} {Erratum: Low-speed impact craters in loose
  granular media [phys. rev. lett.prltao0031-9007 90, 194301 (2003)]},\ }\href
  {https://doi.org/10.1103/PhysRevLett.91.149902} {\bibfield  {journal}
  {\bibinfo  {journal} {Phys. Rev. Lett.}\ }\textbf {\bibinfo {volume} {91}},\
  \bibinfo {pages} {149902} (\bibinfo {year} {2003}{\natexlab{b}})}\BibitemShut
  {NoStop}%
\bibitem [{\citenamefont {Ganapathy}(1980)}]{Ganapathy}%
  \BibitemOpen
  \bibfield  {author} {\bibinfo {author} {\bibfnamefont {R.}~\bibnamefont
  {Ganapathy}},\ }\bibfield  {title} {\bibinfo {title} {A major meteorite
  impact on the earth 65 million years ago: Evidence from the
  cretaceous-tertiary boundary clay},\ }\href
  {https://doi.org/10.1126/science.209.4459.921} {\bibfield  {journal}
  {\bibinfo  {journal} {Science}\ }\textbf {\bibinfo {volume} {209}},\ \bibinfo
  {pages} {921} (\bibinfo {year} {1980})}\BibitemShut {NoStop}%
\bibitem [{\citenamefont {Sawlowicz}(1993)}]{Sawlowicz}%
  \BibitemOpen
  \bibfield  {author} {\bibinfo {author} {\bibfnamefont {Z.}~\bibnamefont
  {Sawlowicz}},\ }\bibfield  {title} {\bibinfo {title} {Iridium and other
  platinum-group elements as geochemical markers in sedimentary environments},\
  }\href {https://doi.org/https://doi.org/10.1016/0031-0182(93)90136-7}
  {\bibfield  {journal} {\bibinfo  {journal} {Palaeogeogr. Palaeocl.}\ }\textbf
  {\bibinfo {volume} {104}},\ \bibinfo {pages} {253} (\bibinfo {year}
  {1993})}\BibitemShut {NoStop}%
\bibitem [{\citenamefont {McDonald}\ \emph {et~al.}(2001)\citenamefont
  {McDonald}, \citenamefont {Andreoli}, \citenamefont {Hart},\ and\
  \citenamefont {Tredoux}}]{McDonald}%
  \BibitemOpen
  \bibfield  {author} {\bibinfo {author} {\bibfnamefont {I.}~\bibnamefont
  {McDonald}}, \bibinfo {author} {\bibfnamefont {M.}~\bibnamefont {Andreoli}},
  \bibinfo {author} {\bibfnamefont {R.}~\bibnamefont {Hart}},\ and\ \bibinfo
  {author} {\bibfnamefont {M.}~\bibnamefont {Tredoux}},\ }\bibfield  {title}
  {\bibinfo {title} {Platinum-group elements in the {M}orokweng impact
  structure, {S}outh {A}frica: {E}vidence for the impact of a large ordinary
  chondrite projectile at the {J}urassic-{C}retaceous boundary},\ }\href
  {https://doi.org/https://doi.org/10.1016/S0016-7037(00)00527-5} {\bibfield
  {journal} {\bibinfo  {journal} {Geochim. Cosmochim. Ac.}\ }\textbf {\bibinfo
  {volume} {65}},\ \bibinfo {pages} {299} (\bibinfo {year} {2001})}\BibitemShut
  {NoStop}%
\bibitem [{\citenamefont {Pacheco-V\'azquez}\ and\ \citenamefont
  {Ruiz-Su\'arez}(2010)}]{Pacheco}%
  \BibitemOpen
  \bibfield  {author} {\bibinfo {author} {\bibfnamefont {F.}~\bibnamefont
  {Pacheco-V\'azquez}}\ and\ \bibinfo {author} {\bibfnamefont {J.}~\bibnamefont
  {Ruiz-Su\'arez}},\ }\bibfield  {title} {\bibinfo {title} {Cooperative
  dynamics in the penetration of a group of intruders in a granular medium},\
  }\href@noop {} {\bibfield  {journal} {\bibinfo  {journal} {Nat. Commun.}\
  }\textbf {\bibinfo {volume} {1}} (\bibinfo {year} {2010})}\BibitemShut
  {NoStop}%
\bibitem [{\citenamefont {Cundall}\ and\ \citenamefont
  {Strack}(1979)}]{Cundall}%
  \BibitemOpen
  \bibfield  {author} {\bibinfo {author} {\bibfnamefont {P.~A.}\ \bibnamefont
  {Cundall}}\ and\ \bibinfo {author} {\bibfnamefont {O.~D.}\ \bibnamefont
  {Strack}},\ }\bibfield  {title} {\bibinfo {title} {A discrete numerical model
  for granular assemblies},\ }\href@noop {} {\bibfield  {journal} {\bibinfo
  {journal} {G\'eotechnique}\ }\textbf {\bibinfo {volume} {29}},\ \bibinfo
  {pages} {47} (\bibinfo {year} {1979})}\BibitemShut {NoStop}%
\bibitem [{\citenamefont {Kloss}\ \emph {et~al.}(2012)\citenamefont {Kloss},
  \citenamefont {Goniva}, \citenamefont {Hager}, \citenamefont {Amberger},\
  and\ \citenamefont {Pirker}}]{Kloss2}%
  \BibitemOpen
  \bibfield  {author} {\bibinfo {author} {\bibfnamefont {C.}~\bibnamefont
  {Kloss}}, \bibinfo {author} {\bibfnamefont {C.}~\bibnamefont {Goniva}},
  \bibinfo {author} {\bibfnamefont {A.}~\bibnamefont {Hager}}, \bibinfo
  {author} {\bibfnamefont {S.}~\bibnamefont {Amberger}},\ and\ \bibinfo
  {author} {\bibfnamefont {S.}~\bibnamefont {Pirker}},\ }\bibfield  {title}
  {\bibinfo {title} {Models, algorithms and validation for opensource dem and
  cfd–dem},\ }\href {https://doi.org/10.1504/PCFD.2012.047457} {\bibfield
  {journal} {\bibinfo  {journal} {Prog. Comput. Fluid Dy.}\ }\textbf {\bibinfo
  {volume} {12}},\ \bibinfo {pages} {140} (\bibinfo {year} {2012})}\BibitemShut
  {NoStop}%
\bibitem [{\citenamefont {Berger}\ \emph {et~al.}(2015)\citenamefont {Berger},
  \citenamefont {Kloss}, \citenamefont {Kohlmeyer},\ and\ \citenamefont
  {Pirker}}]{Berger}%
  \BibitemOpen
  \bibfield  {author} {\bibinfo {author} {\bibfnamefont {R.}~\bibnamefont
  {Berger}}, \bibinfo {author} {\bibfnamefont {C.}~\bibnamefont {Kloss}},
  \bibinfo {author} {\bibfnamefont {A.}~\bibnamefont {Kohlmeyer}},\ and\
  \bibinfo {author} {\bibfnamefont {S.}~\bibnamefont {Pirker}},\ }\bibfield
  {title} {\bibinfo {title} {Hybrid parallelization of the {LIGGGHTS}
  open-source {DEM} code},\ }\href@noop {} {\bibfield  {journal} {\bibinfo
  {journal} {Powder Technol.}\ }\textbf {\bibinfo {volume} {278}},\ \bibinfo
  {pages} {234} (\bibinfo {year} {2015})}\BibitemShut {NoStop}%
\bibitem [{\citenamefont {Di~Renzo}\ and\ \citenamefont
  {Di~Maio}(2004)}]{direnzo}%
  \BibitemOpen
  \bibfield  {author} {\bibinfo {author} {\bibfnamefont {A.}~\bibnamefont
  {Di~Renzo}}\ and\ \bibinfo {author} {\bibfnamefont {F.~P.}\ \bibnamefont
  {Di~Maio}},\ }\bibfield  {title} {\bibinfo {title} {Comparison of
  contact-force models for the simulation of collisions in {DEM}-based granular
  flow codes},\ }\href@noop {} {\bibfield  {journal} {\bibinfo  {journal}
  {Chem. Eng. Sci.}\ }\textbf {\bibinfo {volume} {59}},\ \bibinfo {pages} {525}
  (\bibinfo {year} {2004})}\BibitemShut {NoStop}%
\bibitem [{Sup()}]{Supplemental}%
  \BibitemOpen
  \href@noop {} {\bibinfo  {journal} {See Supplemental Material at [URL to be
  inserted by publisher] for additional graphics for the remaining data, and
  movies showing the motion of grains and the granular temperature as a
  projectile impacts a cohesionless granular bed}\ }\BibitemShut {NoStop}%
\bibitem [{\citenamefont {Guo}\ \emph {et~al.}(2013)\citenamefont {Guo},
  \citenamefont {Wassgren}, \citenamefont {Hancock}, \citenamefont
  {Ketterhagen},\ and\ \citenamefont {Curtis}}]{Guo}%
  \BibitemOpen
\bibfield  {journal} {  }\bibfield  {author} {\bibinfo {author} {\bibfnamefont
  {Y.}~\bibnamefont {Guo}}, \bibinfo {author} {\bibfnamefont {C.}~\bibnamefont
  {Wassgren}}, \bibinfo {author} {\bibfnamefont {B.}~\bibnamefont {Hancock}},
  \bibinfo {author} {\bibfnamefont {W.}~\bibnamefont {Ketterhagen}},\ and\
  \bibinfo {author} {\bibfnamefont {J.}~\bibnamefont {Curtis}},\ }\bibfield
  {title} {\bibinfo {title} {Validation and time step determination of discrete
  element modeling of flexible fibers},\ }\href
  {https://doi.org/https://doi.org/10.1016/j.powtec.2013.09.007} {\bibfield
  {journal} {\bibinfo  {journal} {Powder Technol.}\ }\textbf {\bibinfo {volume}
  {249}},\ \bibinfo {pages} {386} (\bibinfo {year} {2013})}\BibitemShut
  {NoStop}%
\bibitem [{\citenamefont {Schramm}\ \emph {et~al.}(2019)\citenamefont
  {Schramm}, \citenamefont {Tekeste}, \citenamefont {Plouffe},\ and\
  \citenamefont {Harby}}]{Schramm}%
  \BibitemOpen
  \bibfield  {author} {\bibinfo {author} {\bibfnamefont {M.}~\bibnamefont
  {Schramm}}, \bibinfo {author} {\bibfnamefont {M.~Z.}\ \bibnamefont
  {Tekeste}}, \bibinfo {author} {\bibfnamefont {C.}~\bibnamefont {Plouffe}},\
  and\ \bibinfo {author} {\bibfnamefont {D.}~\bibnamefont {Harby}},\ }\bibfield
   {title} {\bibinfo {title} {Estimating bond damping and bond young's modulus
  for a flexible wheat straw discrete element method model},\ }\href
  {https://doi.org/https://doi.org/10.1016/j.biosystemseng.2019.08.003}
  {\bibfield  {journal} {\bibinfo  {journal} {Biosyst. Eng.}\ }\textbf
  {\bibinfo {volume} {186}},\ \bibinfo {pages} {349} (\bibinfo {year}
  {2019})}\BibitemShut {NoStop}%
\bibitem [{\citenamefont {Chen}\ \emph {et~al.}(2022)\citenamefont {Chen},
  \citenamefont {Wang}, \citenamefont {Morrissey},\ and\ \citenamefont
  {Ooi}}]{Chen}%
  \BibitemOpen
  \bibfield  {author} {\bibinfo {author} {\bibfnamefont {X.}~\bibnamefont
  {Chen}}, \bibinfo {author} {\bibfnamefont {L.}~\bibnamefont {Wang}}, \bibinfo
  {author} {\bibfnamefont {J.}~\bibnamefont {Morrissey}},\ and\ \bibinfo
  {author} {\bibfnamefont {J.~Y.}\ \bibnamefont {Ooi}},\ }\bibfield  {title}
  {\bibinfo {title} {{DEM} simulations of agglomerates impact breakage using
  {T}imoshenko beam bond model},\ }\href
  {https://doi.org/10.1007/s10035-022-01231-9} {\bibfield  {journal} {\bibinfo
  {journal} {Granular Matter}\ }\textbf {\bibinfo {volume} {24}} (\bibinfo
  {year} {2022})}\BibitemShut {NoStop}%
\bibitem [{\citenamefont {Gong}\ \emph {et~al.}(2023)\citenamefont {Gong},
  \citenamefont {Yang}, \citenamefont {Cui}, \citenamefont {He},\ and\
  \citenamefont {Liu}}]{Gong}%
  \BibitemOpen
  \bibfield  {author} {\bibinfo {author} {\bibfnamefont {Z.}~\bibnamefont
  {Gong}}, \bibinfo {author} {\bibfnamefont {Y.}~\bibnamefont {Yang}}, \bibinfo
  {author} {\bibfnamefont {L.}~\bibnamefont {Cui}}, \bibinfo {author}
  {\bibfnamefont {J.}~\bibnamefont {He}},\ and\ \bibinfo {author}
  {\bibfnamefont {X.}~\bibnamefont {Liu}},\ }\bibfield  {title} {\bibinfo
  {title} {Dem investigation on the size effect in the fragmentation of intact
  aggregates},\ }\href
  {https://doi.org/https://doi.org/10.1016/j.powtec.2023.118585} {\bibfield
  {journal} {\bibinfo  {journal} {Powder Technol.}\ }\textbf {\bibinfo {volume}
  {425}},\ \bibinfo {pages} {118585} (\bibinfo {year} {2023})}\BibitemShut
  {NoStop}%
\bibitem [{\citenamefont {Ucgul}\ \emph
  {et~al.}(2014{\natexlab{a}})\citenamefont {Ucgul}, \citenamefont {Fielke},\
  and\ \citenamefont {Saunders}}]{Ucgul1}%
  \BibitemOpen
  \bibfield  {author} {\bibinfo {author} {\bibfnamefont {M.}~\bibnamefont
  {Ucgul}}, \bibinfo {author} {\bibfnamefont {J.~M.}\ \bibnamefont {Fielke}},\
  and\ \bibinfo {author} {\bibfnamefont {C.}~\bibnamefont {Saunders}},\
  }\bibfield  {title} {\bibinfo {title} {{3D DEM} tillage simulation:
  Validation of a hysteretic spring (plastic) contact model for a sweep tool
  operating in a cohesionless soil},\ }\href@noop {} {\bibfield  {journal}
  {\bibinfo  {journal} {Soil Till. Res.}\ }\textbf {\bibinfo {volume} {144}},\
  \bibinfo {pages} {220} (\bibinfo {year} {2014}{\natexlab{a}})}\BibitemShut
  {NoStop}%
\bibitem [{\citenamefont {Ucgul}\ \emph
  {et~al.}(2014{\natexlab{b}})\citenamefont {Ucgul}, \citenamefont {Fielke},\
  and\ \citenamefont {Saunders}}]{Ucgul2}%
  \BibitemOpen
  \bibfield  {author} {\bibinfo {author} {\bibfnamefont {M.}~\bibnamefont
  {Ucgul}}, \bibinfo {author} {\bibfnamefont {J.~M.}\ \bibnamefont {Fielke}},\
  and\ \bibinfo {author} {\bibfnamefont {C.}~\bibnamefont {Saunders}},\
  }\bibfield  {title} {\bibinfo {title} {Three-dimensional discrete element
  modelling of tillage: Determination of a suitable contact model and
  parameters for a cohesionless soil},\ }\href@noop {} {\bibfield  {journal}
  {\bibinfo  {journal} {Biosyst. Eng.}\ }\textbf {\bibinfo {volume} {121}},\
  \bibinfo {pages} {105} (\bibinfo {year} {2014}{\natexlab{b}})}\BibitemShut
  {NoStop}%
\bibitem [{\citenamefont {Ucgul}\ \emph {et~al.}(2015)\citenamefont {Ucgul},
  \citenamefont {Fielke},\ and\ \citenamefont {Saunders}}]{Ucgul3}%
  \BibitemOpen
  \bibfield  {author} {\bibinfo {author} {\bibfnamefont {M.}~\bibnamefont
  {Ucgul}}, \bibinfo {author} {\bibfnamefont {J.~M.}\ \bibnamefont {Fielke}},\
  and\ \bibinfo {author} {\bibfnamefont {C.}~\bibnamefont {Saunders}},\
  }\bibfield  {title} {\bibinfo {title} {Three-dimensional discrete element
  modelling (dem) of tillage: Accounting for soil cohesion and adhesion},\
  }\href@noop {} {\bibfield  {journal} {\bibinfo  {journal} {Biosyst. Eng.}\
  }\textbf {\bibinfo {volume} {129}},\ \bibinfo {pages} {298} (\bibinfo {year}
  {2015})}\BibitemShut {NoStop}%
\bibitem [{\citenamefont {Derakhshani}\ \emph {et~al.}(2015)\citenamefont
  {Derakhshani}, \citenamefont {Schott},\ and\ \citenamefont
  {Lodewijks}}]{Derakhshani}%
  \BibitemOpen
  \bibfield  {author} {\bibinfo {author} {\bibfnamefont {S.~M.}\ \bibnamefont
  {Derakhshani}}, \bibinfo {author} {\bibfnamefont {D.~L.}\ \bibnamefont
  {Schott}},\ and\ \bibinfo {author} {\bibfnamefont {G.}~\bibnamefont
  {Lodewijks}},\ }\bibfield  {title} {\bibinfo {title} {Micro–macro
  properties of quartz sand: {E}xperimental investigation and {DEM}
  simulation},\ }\href
  {https://doi.org/https://doi.org/10.1016/j.powtec.2014.08.072} {\bibfield
  {journal} {\bibinfo  {journal} {Powder Technol.}\ }\textbf {\bibinfo {volume}
  {269}},\ \bibinfo {pages} {127} (\bibinfo {year} {2015})}\BibitemShut
  {NoStop}%
\bibitem [{\citenamefont {Zaikin}\ \emph {et~al.}(2017)\citenamefont {Zaikin},
  \citenamefont {Korablin}, \citenamefont {Dyulger},\ and\ \citenamefont
  {Barnenkov}}]{Zaikin}%
  \BibitemOpen
  \bibfield  {author} {\bibinfo {author} {\bibfnamefont {O.}~\bibnamefont
  {Zaikin}}, \bibinfo {author} {\bibfnamefont {A.}~\bibnamefont {Korablin}},
  \bibinfo {author} {\bibfnamefont {N.}~\bibnamefont {Dyulger}},\ and\ \bibinfo
  {author} {\bibfnamefont {N.}~\bibnamefont {Barnenkov}},\ }\bibfield  {title}
  {\bibinfo {title} {Model of the relationship between the velocity restitution
  coefficient and the initial car velocity during collision},\ }\href@noop {}
  {\bibfield  {journal} {\bibinfo  {journal} {Transp. Res. Proc.}\ }\textbf
  {\bibinfo {volume} {20}},\ \bibinfo {pages} {717} (\bibinfo {year}
  {2017})}\BibitemShut {NoStop}%
\bibitem [{\citenamefont {Lima}\ \emph {et~al.}(2023)\citenamefont {Lima},
  \citenamefont {Carvalho},\ and\ \citenamefont {Franklin}}]{Supplemental2}%
  \BibitemOpen
  \bibfield  {author} {\bibinfo {author} {\bibfnamefont {N.~C.}\ \bibnamefont
  {Lima}}, \bibinfo {author} {\bibfnamefont {D.~D.}\ \bibnamefont {Carvalho}},\
  and\ \bibinfo {author} {\bibfnamefont {E.~M.}\ \bibnamefont {Franklin}},\
  }\href@noop {} {\bibfield  {journal} {\bibinfo  {journal} {{LIGGGHTS} input
  and output files, and Octave scripts for post-processing the outputs are
  available on Mendeley Data, http://dx.doi.org/10.17632/d49b9p6f4r.1}\ }
  (\bibinfo {year} {2023})}\BibitemShut {NoStop}%
\end{thebibliography}%

\end{document}